\documentclass[hyper, letterpaper,11pt,notoc]{JHEP3}
\usepackage{graphicx} 
\usepackage{amssymb}
\usepackage{amsmath} 
\usepackage{slashed}
\newcommand{\GeV}{~\mathrm{GeV}}
\newcommand{\TeV}{~\mathrm{TeV}}

\title{A T-odd observable sensitive to CP violating phases in squark decay}
\author{Paul Langacker$^{(a)}$, Gil Paz$^{(a)}$, Lian-Tao Wang$^{(b)}$ and Itay Yavin$^{(b)}$ \\ \it{(a) School of Natural Sciences, Institute for Advanced Study, Einstein Drive Princeton, NJ 08540} \\ \it{(b) Department of Physics, Princeton University, Princeton NJ.
08544}}
\abstract{We present a new observable sensitive to a certain combination of CP violating phases in supersymmetric extensions of the Standard Model, viz. a triple product of momenta in the cascade decay of
a heavy squark via an on-shell neutralino and off-shell slepton. We investigate the regions of parameter space in which the signal is strong enough to be detectable at the LHC with $\sim \bigl(10^2-10^3\bigr)/\sin^2(2\Delta\varphi)$ identified
events, where $\Delta\varphi$ is a certain combination of phases in the MSSM presented in the text.}

\preprint{} 
\begin{document}
\section{Introduction}
In the past decade it became clear that the Standard Model cannot account for electroweak baryogenesis (for
recent reviews, see \cite{Trodden:1998ym,Riotto:1999yt,Bernreuther:2002uj,Dine:2003ax}). The CP violating phases (the CKM phase and the strong phase) are too small, and with the Higgs mass constrained by LEP2 to be $m_{H} > 115\GeV$ the electroweak phase transition is not strong enough to suppress the sphaleron process. 

However, it is still possible that baryogenesis is connected with electroweak scale physics and has to do with the other missing ingredient from the Standard Model, which is the mechanism for stabilizing the electroweak scale. If that mechanism is supersymmetry, then many new possibilities open up. The new supersymmetric sectors contain, in principle, many new phases. Also, there are many more contributions to the Higgs potential, and the electroweak phase transition can be made stronger. 

In the context of the MSSM, there is only a small window in parameter space left to accommodate electroweak baryogenesis \cite{Carena:1996wj,Carena:1997ki,Carena:2000id,Carena:2002ss}. One difficulty arises 
because there are no cubic terms in the tree-level Higgs potential, and the bounds from LEP2 on the Higgs and stop masses severely restrict the loop-induced contributions   and disfavor a strong first-order phase transition. Nonetheless, 
there seems to be no difficulty in generating a strong first order transition in slightly more general supersymmetric models,
such as those involving extra singlet Higgs in which there can be cubic terms at
tree-level  
\cite{Pietroni:1992in,Davies:1996qn,Huber:1998ck,Huber:2000mg,Kang:2004pp,Menon:2004wv,Huber:2006wf}. 
The other difficulty is in ensuring sufficiently large CP violating phases, given the stringent bounds
on certain combinations of such phases from electron and neutron  electric dipole moment (EDM)  experiments \cite{Pokorski:1999hz,Ibrahim:1999af,Barger:2001nu,Pospelov:2005pr}. Since the operator responsible for EDM involves a helicity flip, these bounds mostly constrain the phase combination involving the $\mu$-parameter and the wino $M_2$ phases. The same phase combination is responsible for the dominant electroweak baryogenesis mechanisms in the MSSM, leading to tension with the EDM constraints. This may also be relaxed in extended models involving additional phases \cite{Pietroni:1992in}-\cite{Huber:2006wf},\cite{Demir:2003ke}.

Motivated by the possible abundance of CP violating phases in supersymmetric extensions of the Standard Model \cite{Brhlik:1998gu},
their relevance to baryogenesis, and their impact on Higgs and sparticle spectra \cite{Mrenna:1999ai,Kane:2000aq}
 it is important to explore them in a variety of ways.
In particular, there have been a number of suggestions for their direct detection in collider experiments 
\cite{Donoghue:1987ax,Im:1993ur,Dawson:1995wg,Arkani-Hamed:1997km,Nachtmann:2003hg,Bartl:2004jr,Valencia:2005cx,Kiers:2006aq,Szynkman:2007uc}. 

The MSSM soft lagrangian contains many different phases, however, not all are physical. As emphasized in \cite{Chung:2003fi}, the physical phases are those which are invariant under both $U_R(1)$ and $U_{PQ}(1)$ symmetries. There are several obvious such invariant: (i) the phases in the off diagonal elements of the soft scalar masses; (ii) the relative phases between the gaugino masses; (iii) the relative phases between the different A parameters. The phases which are affected by the $R$ and $PQ$ rotations are $\phi_\mu$, $\phi_b$ (the phases of the $\mu$ and $b$ terms), $\phi_{M_a}$ and $\phi_{\tilde{A}_f}$, the overall phases of the $A$ parameters. One can then build reparameterization invariant linear combinations of these phases which appear in physical processes (for one such parameterization see \cite{Lebedev:2002wq}) 

EDM experiments are sensitive to linear combinations of phases involving $\phi_\mu$ since the dominant diagram involves the chargino exchange (see for example \cite{Barger:2001nu}). In this paper we present a new observable which is sensitive to a different combination of the phases than is probed
in the EDM experiments. It involves the phases in the couplings of the neutralinos to the sleptons,
and in a typical limiting case depends on the difference between the phases of the bino and wino
mass parameters $M_1$ and $M_2$ (which can differ for nonuniversal gaugino masses). Since it requires no higgsino insertion it is insensitive to the phase of the $\mu$-parameter. 

We propose an asymmetry parameter related to the usual, T-violating, triple product $\langle \vec{p}_1\cdot (\vec{p}_2\times \vec{p}_3)\rangle$, with $\vec{p}_i$ being 3 independent vectors in a given reaction. This quantity constitutes a direct measurement of T-invariance violation, which, in the context of CPT invariant theories, is a measurement of CP-violation. 

As can be expected on general grounds, any such asymmetry parameter is the result of interfering diagrams. As such, it is usually suppressed with respect to the leading non-interfering contribution. In the case of a reaction which proceeds through an on-shell cascade decay,  interference terms are suppressed with respect to the on-shell amplitudes by a factor of the width. However, if some part of the reaction is forced to proceed off-shell, the interference terms are comparable to the leading amplitudes. We exploit this fact in the decay of a squark into a quark and two leptons $\tilde{q} \rightarrow q + \tilde{N}_2 \rightarrow q + l^+ + l^- + \tilde{N}_1$,
where $\tilde{N}_{1,2}$ are neutralinos and we assume that $\tilde N_1$ is the lightest
supersymmetric particle (LSP). We show that in regions of parameter space where the decay of $\tilde{N}_{2}$ is through an off-shell slepton the asymmetry parameter, $\eta$ (defined below), can be very large, $\eta \sim \mathcal{O}(1)$. 

The number of events required for  determining  $\eta$ scales as $N \propto 1/\eta^2$ (assuming Gaussian statistics). This calls for as precise a theoretical estimate of $\eta$ as possible, since a factor of $3$ reduction can translate into an order of magnitude more events. Clearly, the actual number of events needed is affected by many experimental considerations as well. Reduction of the signal due to mis-tagging, detector resolution, tagging efficiency, etc., will inevitably increase the number of events required. After presenting the theoretical results we attempt to estimate the signal reduction due to experimental limitations and present the number of events needed to determine the existence of a CP asymmetry.

The paper is organized as follows: In section \ref{sec:GeneralDiscussion} we present a general discussion of  the T, CP,
and P-violating triple product and the relevant observables. In section \ref{sec:on-shell} we consider the reaction $\tilde{t} \rightarrow t + l^+ + l^- + \tilde{N}_1$ and show that it contains appropriate observables sensitive to certain CP-phases in the MSSM. However, we find that if the reaction proceeds through an on-shell cascade decay, the signal is too small to be measured. It is also possible for the reaction to proceed via an off-shell cascade. In that case, as we shall see in section \ref{sec:off-shell}, the effect is greatly enhanced and the signal may be large enough to be detectable. In section \ref{sec:ExpLim}  we attempt to estimate the reduction in the signal strength due to experimental limitations and section \ref{sec:conclusions} contains our conclusions.

\section{General discussion}
\label{sec:GeneralDiscussion}
In this paper we will concentrate on CP-violating effects present in cascade decays of heavy supersymmetric particles. When an unstable particle decays through a reaction involving 4 other particles, it is possible that the expression for the rate contains a contribution of the form\footnote{Our conventions are $g_{\mu\nu}={\rm diag} (1,-1,-1,-1)$
and $\epsilon_{0123}=-1$.}
$\epsilon_{\mu\nu\alpha\beta} p_0^\mu p_1^\nu p_2^\alpha p_3^\beta$, where $p_i$ are four independent momenta. In the rest frame of the decaying particle, $p_0=(M_0,0,0,0)$, this term is the usual P and T-odd observable given by the expectation value of the triple product
\begin{equation}
\label{eqn:TripleProduct}
{\cal T} =-M_0 ~\vec{p}_1 \cdot \left(\vec{p}_2 \times \vec{p}_3\right),
\end{equation}
where $M_0$ is the mass of the decaying particle.
As is well known \cite{Gasiorowicz}, any measurement of a non-zero expectation value
$\langle \cal T \rangle$ implies both P violation and  either T violation {\em or} a ``strong phase''.
Assuming that CPT invariance is unbroken, T violation is equivalent to the non-conservation of CP,
and is manifested by a CP-violating phase in the Lagrangian. A strong phase refers to a CP-conserving
phase, due, e.g., to a strong or electromagnetic final state interaction or the phase associated
with the width in the propagator of an unstable particle. The two effects can in principle
be separated by measuring the expectation values of both $\cal T$ and $\overline {\cal T}$,
where
$\overline {\cal T} = -M_0\vec{p}_1\!^c \cdot \left(\vec{p}_2\!^c \times \vec{p}_3\!^c\right)$
is the corresponding triple product for the decay of the antiparticle, and $p_i\!^c$ are the
physical momenta of the antiparticles in the final state.
Since $\cal T$ is both P and T-odd a non-zero $\langle \cal T \rangle$ requires the
interference of two contributions to the amplitude involving different phases and different
parity, i.e., the amplitude in the rest frame $\vec p_0=0$ must contain
\begin{equation}
\label{eqn:two_amps}
\langle \vec p_i | H | \vec p_0\rangle =A(\vec p_i) e^{i(\rho_A + \phi_A)}
+ B(\vec p_i) e^{i(\frac{\pi}{2}+\rho_B + \phi_B)},
\end{equation}
where $\vec p_i$ refers collectively to the final momenta, and,
\begin{align}
\label{eqn:ABparity}
A(-\vec p)&=A(\vec p) \\\nonumber
B(-\vec p)&=-B(\vec p).  
\end{align}
$\phi_{A,B}$ and $\rho_{A,B}$ are respectively the so-called
weak and strong  phases,  which do (do not) change sign in the CP-conjugate
process\footnote{For a more thorough discussion of these phases see the excellent textbooks
\cite{Branco:1999fs,Bigi:2000yz}.}. The phase $\pi/2$ in the second term could have been
absorbed in $\rho_B$, but is instead pulled out for convenience. It always
occurs in the relevant interference term between the parity even and odd
amplitudes for  $\langle \cal T \rangle$  when one sums over the spins
\cite{Gasiorowicz}. For the example considered in this paper, it is just the
explicit factor of $i$ occurring in the trace of $\gamma^5 \gamma_\mu \gamma_\nu
\gamma_\alpha \gamma_\beta$. Using (\ref{eqn:two_amps}), it is then straightforward
to show that
\begin{equation}
\label{eqn:exp_value}
\langle{ \cal T} \rangle = 2 {\cal K} \left[ \sin(\rho_A-\rho_B) \cos(\phi_A-\phi_B)+
\cos(\rho_A-\rho_B) \sin(\phi_A-\phi_B) \right],
\end{equation}
where
${\cal K} $ is proportional to the phase space integral $ \int dPS({\cal T} A B)$. The
first (second) term in (\ref{eqn:exp_value}) requires a nonzero difference between
the strong (weak) phases. Using CP, the corresponding amplitude for the
antiparticle decay is
  \begin{equation}
\label{eqn:conj_amps}
\langle \vec p_i\!^c | H | \vec p_0\!^c \rangle =A(-\vec p_i\!^c) e^{i(\rho_A - \phi_A)}
+ B(-\vec p_i\!^c) e^{i(\frac{\pi}{2}+\rho_B - \phi_B)},
\end{equation}
up to an irrelevant overall phase associated with our CP conventions. Comparing with (\ref{eqn:two_amps})
and using the parities of $A$ and $B$ in (\ref{eqn:ABparity}), one sees
that $\langle\overline{ \cal T} \rangle$ differs from $\langle \cal T \rangle$
by an overall sign (due to the fact that the observable is P-odd) and by
$\phi_{A,B} \rightarrow - \phi_{A,B} $,
\begin{equation}
\label{eqn:conj_value}
\langle{\overline{ \cal T}} \rangle = -2 {\cal K} \left[\sin(\rho_A-\rho_B) \cos(\phi_A-\phi_B)-
\cos(\rho_A-\rho_B) \sin(\phi_A-\phi_B) \right].
\end{equation}
In particular, the T-odd term can be isolated by summing the particle and
antiparticle asymmetries
\begin{equation}
\label{eqn:sum_value}
\langle{ \cal T} \rangle +\langle{\overline{\cal T}} \rangle= 4 {\cal K}
\cos(\rho_A-\rho_B) \sin(\phi_A-\phi_B).
\end{equation}

In this paper we will present a manifestation of these general considerations in the particular supersymmetric cascade decays
\begin{equation}
\tilde{t} \rightarrow t + \tilde{N}_a \rightarrow t + l^+ + l^- + \tilde{N}_1,
\end{equation}
and their CP conjugates
$\tilde{t}^c \rightarrow t^c + \tilde{N}_a \rightarrow t^c + l^- + l^+ + \tilde{N}_1$,
where $\tilde{N}_a$ (which we usually take to be $\tilde N_2$) is assumed to be on-shell.
These reactions have enough independent momenta and, as we shall show below, lead to  non-vanishing expectation values for the triple product, Eq. (\ref{eqn:TripleProduct}), made of the top and di-lepton momenta.  $\cal T$ and $\overline{\cal T}$
refer respectively to the observables $-M_{\tilde{t}}~\vec{p}_t \cdot \left(\vec{p}_{l^+} \times \vec{p}_{l^-}\right)$
and $-M_{\tilde{t}}~\vec{p}_{t^c} \cdot \left(\vec{p}_{l^-} \times \vec{p}_{l^+}\right)$.

\begin{figure}
\begin{center}
\includegraphics[scale=0.7]{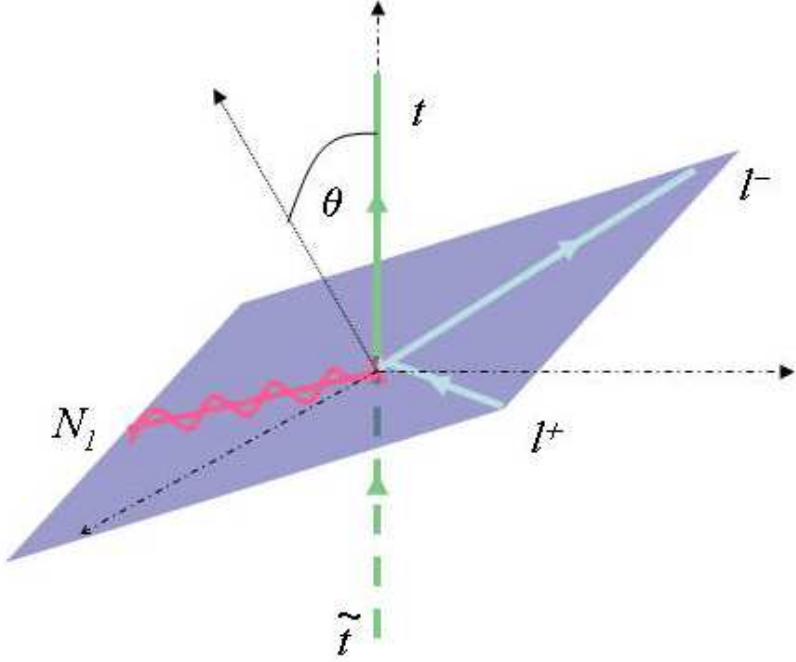}
\end{center}
\caption{The reaction geometry in the rest frame of $\tilde{N}_a$
for $\tilde{t}  \rightarrow t + l^+ + l^- + \tilde{N}_1$.
$\theta$ is the angle between $\vec p_t$ and $ \vec{p}_{l^+} \times \vec{p}_{l^-}$.
For $\tilde{t}^c \rightarrow t^c  + l^- + l^+ + \tilde{N}_1$,
$\theta$ is the angle between $ \vec{p}_{t^c}$ and $ \vec{p}_{l^-} \times \vec{p}_{l^+}$.}
\label{fig:z-plane-angle}
\end{figure}

It proves useful to formulate the discussion in terms of a dimensionless parameter embodying the CP asymmetry. This parameter is closely related to the triple product with the added advantage of allowing for a straightforward evaluation of the number of events needed. As shown in Fig. \ref{fig:z-plane-angle}, in the rest frame of the on-shell $\tilde{N}_a$, the incoming $\tilde{t}$ and outgoing top define a $z$-axis. Momentum conservation forces the outgoing anti-lepton, lepton and the LSP to define a plane. A non-zero expectation value of $\cal T$ implies a non-zero average angle between the plane and the $z$-axis. 
 We therefore define the asymmetry parameter
\begin{equation}
\label{eqn:eta_def}
\eta = \frac{N_+ - N_-}{N_+ + N_-}  = \frac{N_+ - N_-}{N_{total}},
\end{equation}
where
\begin{equation}
\label{eqn:N+N-}
N_+ = \int_0^{1} \frac{d\Gamma}{d\cos\theta} d\cos\theta,  \qquad N_- = \int_{-1}^{0} \frac{d\Gamma}{d\cos\theta} d\cos\theta.
\end{equation}

Below, we derive an exact expression for  $\eta$ in the neutralino's rest frame. Clearly, $\eta$ is not a relativistically invariant variable. If the LSP escapes detection it is impossible to reconstruct the neutralino's rest frame. Therefore, $\eta $ can only be constructed in the lab frame (i.e., detector frame), which inevitably affects the signal. In section \ref{sec:ExpLim} we show that this lack of knowledge of the correct frame gives rise to a dilution factor $\mathcal{D}$, similar to that encountered in $B$ physics experiments,
\begin{equation}
\eta_{exp} = \mathcal{D} \eta_{th}.
\end{equation}

Assuming Gaussian statistics, the number of events needed to make a statistically significant measurement is given by
\begin{equation}
 \label{eqn:nstat}
N= \frac{1}{\eta_{exp}^2} = \frac{1}{\mathcal{D}^2 \eta_{th}^2}.
\end{equation}
The dilution factor, as its name implies, leads to an increase in the number of events needed. In what follows we evaluate the asymmetry parameter $\eta_{th}$ and show that in certain kinematical regimes it can
be rather large $\mathcal{O}(1)$. Also, we find that the dilution factor $\mathcal{D}$ need not be very small and in fact does not present a very serious obstacle.

\section{CP-violation in the stop cascade decay via an on-shell slepton}
\label{sec:on-shell}

In this section we compute the expectation value of $\epsilon_{\mu\nu\alpha\beta} p_{\tilde{t}}^\mu p_{t}^\nu p_{l^+}^\alpha p_{l^-}^\beta$ for the cascade decay $\tilde{t}  \rightarrow t + l^+ + l^- + \tilde{N}_1$ via the two diagrams shown in Fig. \ref{fig:Mab} and written explicitly in the appendix.

\begin{figure}[h]
\begin{center}
\includegraphics[scale=0.5]{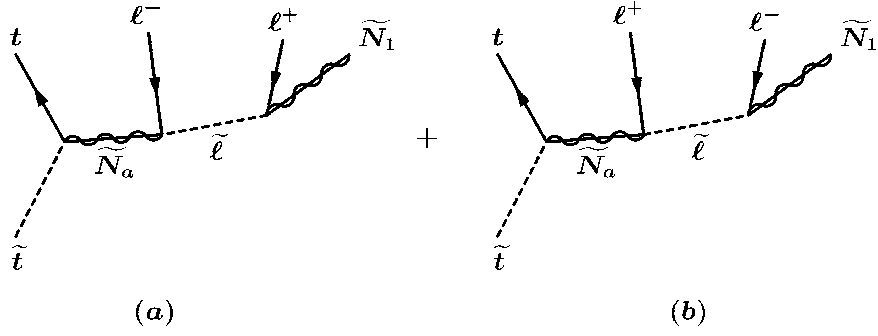}
\end{center}
\caption{The Feynman diagrams corresponding to the matrix elements $i\mathcal{M}_a$ (left) and $i\mathcal{M}_b$ (right). The full matrix element is $i\mathcal{M} = i\mathcal{M}_a + i\mathcal{M}_b$.} 
\label{fig:Mab}
\end{figure}
The relativistically invariant expectation value is given by
\begin{equation}
\label{eqn:EVdef}
\langle\epsilon_{\mu\nu\alpha\beta}~ p_{\tilde{t}}^\mu ~p_{t}^\nu p_{l^+}^\alpha ~ p_{l^-}^\beta ~\rangle =\frac{ \int d\Gamma  \epsilon_{\mu\nu\alpha\beta}~ p_{\tilde{t}}^\mu ~p_{t}^\nu~ p_{l^+}^\alpha~ p_{l^-}^\beta~}{\int d\Gamma},
\end{equation}
where $p_{\tilde{t}}$, $p_{t}$, $p_{l^+}$ and $p_{l^-}$ are the four-momenta of the squark, quark, anti-lepton and lepton, respectively, and $d\Gamma$ is the differential decay width.

Assuming that the neutralino $\tilde{N}_a$ is on-shell the expectation value can be evaluated exactly; the details are presented in the appendix. Considering only flavor diagonal interactions, the numerator in Eq. (\ref{eqn:EVdef}) is
\begin{align}
\int d\Gamma  \epsilon_{\mu\nu\alpha\beta}~ p_{\tilde{t}}^\mu ~p_{t}^\nu~ p_{l^+}^\alpha~ p_{l^-}^\beta~ &= \frac{1}{3} \frac{M_{\tilde{N}_a}^4 ~|\vec{p}_t|^2}{256\pi^3} \left(\frac{M_{\tilde{N}_a}}{\Gamma_{\tilde{N}_a}}\right) \left( \int\frac{dPS_2}{2M_{\tilde{t}}} \right) 
\\\nonumber &\times\left(|g_L^{qa}|^2 - |g_R^{qa}|^2\right)\\\nonumber &\times \left(2~\textrm{Im}\left(g_R^{la*}g_L^{la*}g_R^{l1}g_L^{l1} \right)+\frac{M_{\tilde{N}_1}}{M_{\tilde{N}_a}}  \textrm{Im}\left[ \left(g_R^{la*}\right)^2 ~\left(g_R^{l1}\right)^2 +  R\leftrightarrow L \right]\right)  
\\\nonumber &\times \int dx_+ dx_- f(x_+,x_-),
\end{align}
where
\begin{equation}
|\vec{p}_t|^2 = \frac{1}{4M_{\tilde{N}_a}^2} \left( (M_{\tilde{t}}^2-M_{\tilde{N}_a}^2-M_t^2)^2 - 4M_{\tilde{N}_a}^2 M_t^2 \right).
\end{equation}
The dimensionless function $f(x_+,x_-)$ is given by
\begin{align}
f(x_+,x_-) &= \left( 1 -\mu_1-x_+ - x_-+x_+x_-\right)\left(x_+ + x_-+\mu_1-1 \right) \\\nonumber &\times   \frac{\left((1-\mu_{\tilde{l}} -x_+)(1-\mu_{\tilde{l}} -x_-) + \mu_{\tilde{l}}^2 \gamma_{\tilde{l}}  \right) }{\left((1-\mu_{\tilde{l}} -x_+)^2 +  \mu_{\tilde{l}}^2 \gamma_{\tilde{l}}\right)\left((1-\mu_{\tilde{l}} -x_-)^2 +  \mu_{\tilde{l}}^2 \gamma_{\tilde{l}}\right)  },
\end{align}
where
\begin{equation}
\mu_{\tilde{l}} = \frac{M_{\tilde{l}}^2}{M_{\tilde{N}_a}^2} , \quad \mu_1 = \frac{M_{\tilde{N}_1}^2}{M_{\tilde{N}_a}^2} \quad  \text{and} \quad \gamma_{\tilde{l}} = \frac{\Gamma_{\tilde{l}}^2}{M_{\tilde{l}}^2}.
\end{equation}

To arrive at an expression for the expectation value we need the total width of the reaction. In the narrow-width limit the computation is straightforward and can be carried out analytically. The result is given by
\begin{eqnarray}
\label{eqn:totalW}
\int d\Gamma &=& \frac{M_{\tilde{N}_a}^2}{256\pi^2}\frac{1}{\Gamma_{\tilde{l}} M_{\tilde{l}}}  \left( \frac{ M_{\tilde{N}_a}}{\Gamma_{\tilde{N}_a}}\right)\left(\int\frac{dPS_2}{2M_{\tilde{t}}}\right)
~~ \frac{1}{\mu_{\tilde{l}}} \left(\mu_{\tilde{l}}-\mu_1\right)^2 ~\left(1-\mu_{\tilde{l}}\right)^2
\\\nonumber
\\\nonumber &\times&  ~ \left(|g_L^{l1}|^2 + |g_R^{l1}|^2 \right) ~\left(|g_L^{la}|^2 + |g_R^{la}|^2  \right) 
\\\nonumber
\\\nonumber  &\times& \left( (M_{\tilde{t}}^2-M_{\tilde{N}_a}^2-M_t^2)\left(|g_L^{qa}|^2 + |g_R^{qa}|^2\right) + 4 M_t M_{\tilde{N}_a} \textrm{Re}(g_L^{qa} g_R^{qa*})\right).
\end{eqnarray} 

Clearly, the general expression for the expectation value is fairly complicated. It is instructive to consider some limits where it simplifies. For example, if we assume there is no suppression from the kinematical factors (e.g. $(1-\mu_{\tilde{l}})\sim1$) in Eq. (\ref{eqn:totalW}), and considering the limit where $\tilde{N}_a$ is a pure wino and $\tilde{N}_1$ a pure bino, the final expression is,
\begin{eqnarray}
\label{eqn:EpsilonRes}
\langle\epsilon_{\mu\nu\alpha\beta}~ p_{\tilde{t}}^\mu ~p_{t}^\nu~ p_{l^+}^\alpha ~ p_{l^-}^\beta ~\rangle &=& \frac{1}{24}M_{\tilde{t}}^2M_{\tilde{N}_a}^2~\left(\frac{1}{\pi}\frac{\Gamma_{\tilde{l}}}{M_{\tilde{l}}}\right) \sqrt{\mu_1}  \int dx_+ dx_- f(x_+,x_-) 
\\\nonumber\\\nonumber &\times& \sin\left(2\varphi_L^{l1}-2\varphi_L^{la}\right)
\end{eqnarray}
where we have written the complex couplings as $g_L^{l1}=|g_L^{l1}|\exp(i\varphi_L^{l1})$, etc. This expression is suppressed by the width of the slepton and one additional phase-space factor. This is simply a reflection of the fact that the signal is the ratio of an off-shell process to an on-shell one, i.e., the two diagrams in
Fig. \ref{fig:Mab} cannot simultaneously be on-shell except for a set of measure zero. 
The integral over $f(x_+,x_-)$ is of order unity and cannot enhance the signal. If the LSP is mostly a bino then the slepton decay width is roughly
\begin{equation}
\label{eqn:sleptonWidth}
\frac{1}{\pi} \frac{\Gamma_{\tilde{l}}}{M_{\tilde{l}} }\sim \frac{\alpha_e}{\pi} \sim \frac{1}{300}.
\end{equation}
In order to reliably estimate the number of events needed to reach experimental sensitivity one must form a dimensionless quantity, such as the asymmetry variable presented in the previous section, Eq. (\ref{eqn:eta_def}). The dimensionful phase-space factors in Eq.(\ref{eqn:EpsilonRes}) roughly cancel out in such an observable. Therefore we expect that the quantity in (\ref{eqn:sleptonWidth}) gives us a good order of magnitude estimate for the number of events needed. Without even taking experimental limitations into account we need at least $10^5$ events to reach statistical significance. 

However, if one is close to the decay threshold such that $M_{\tilde{N}_a}^2-M_{\tilde{l}}^2 \lesssim M_{\tilde{l}}\Gamma_{\tilde{l}}$,  the decay rate is suppressed and the asymmetry is enhanced. A more likely possibility is a spectrum where the slepton is forced to be off-shell. In this case there is no width suppression. We explore this possibility in the next section and show that indeed the signal is greatly enhanced.

\section{Stop cascade decay via an off-shell slepton}
\label{sec:off-shell}
In this section we consider the case where $M_{\tilde{l}}>M_{\tilde{N}_a}$. The neutralino may decay through the 3-body channel $\tilde{N}_a\rightarrow l^+ + l^- + \tilde{N}_1$ via an off-shell slepton. 
While most of the results hold for general $\tilde{N}_a$, to simplify the discussion we will often take $\tilde{N}_a$ to be approximately wino.
In general, there may be additional decay paths to consider, and in particular the neutralino can decay directly into $\tilde{N}_1$ and a $Z$-boson. If $M_{\tilde{N}_a}-M_{\tilde{N}_1}>M_Z$ then the $Z$ is on-shell and this channel dominates over the 3-body mode. However, if $M_{\tilde{N}_a}-M_{\tilde{N}_1}< M_Z$ then the $Z$ is off-shell and this reaction might compete with the diagram involving an off-shell slepton. Which is dominant is a detailed question depending on the spectrum. 
The coupling $ \bar{\tilde{N}}_a \tilde{N}_1 Z$ is a result of mixing with the higgsino component. For $a=2$ it is therefore governed by the size of the $\mu$ term compared 
with the gaugino masses $M_1$ and $M_2$ and  is decoupled in the limit of large $\mu$.
Our major goal is to illustrate the possibility of measuring a CP-violating effect rather than to exhaustively 
examine all of parameter space.
In what follows we will therefore ignore the possible contribution of the $ \bar{\tilde{N}}_a \tilde{N}_1 Z$  vertex. In the more general case the interference
with the $Z$ diagram could enhance or reduce the effect.

We expect the signal to have no parametric suppression as in the case of an on-shell decay discussed in the previous section.  Also,  to evaluate the number of events needed we concentrate on the asymmetry parameter $\eta$ defined in Eq. (\ref{eqn:eta_def}). The details are very similar to the previous section except that the interference terms in the width cannot be neglected. Therefore, for integrated luminosity $\mathcal{L}$ the total number of events is given by
\begin{equation}
\label{eqn:Total}
\frac{N_{\rm{total}}}{\mathcal{L}} = \int \frac{dPS_4}{2M_{\tilde{t}}} \left(|\mathcal{M}_a|^2 + |\mathcal{M}_b|^2 + 2\textrm{Re}(\mathcal{M}_a\mathcal{M}_b^*) \right).
\end{equation}

The evaluation of the phase-space integrals is presented in the appendix. The difference between the number of events in the upper and lower hemispheres is given by
\begin{align}
\label{eqn:Difference}
&\frac{N_+ - N_-}{\mathcal{L}} =  \frac{1}{256\pi^3}\left(\frac{M_{\tilde{N}_a}}{\Gamma_{\tilde{N}_a}}\right) \left(\int\frac{dPS_2}{2M_{\tilde{t}}} \right)\left(\frac{|\vec{p}_t|}{M_{\tilde{N}_a}}\right)\left(|g_R^{q}|^2 - |g_L^{q}|^2 \right)\\\nonumber &\times \left(  M_{\tilde{N}_a} M_{\tilde{N}_1}  \textrm{Im}\left[ \left(g_R^{la*}\right)^2 \left(g_R^{l1}\right)^2 + R\leftrightarrow L\right]+ 2M_{\tilde{N}_a}^2\textrm{Im}\left( g_R^{la*}g_L^{la*}g_R^{l1}g_L^{l1}\right) \right) \\\nonumber &\times \int dx_+ dx_- \frac{\Bigl(\left(1-\mu_1-x_+ - x_-+x_+x_-\right)\left(x_+ + x_- +\mu_1-1 \right)\Bigr)^{1/2}} {\left(1-x_+-\mu_{\tilde{l}}\right)\left(1-x_--\mu_{\tilde{l}}\right)}.
\end{align}
The integrals evaluate to
\begin{align}
\label{eqn:notIntTerms}
\int \frac{dPS_4}{2M_{\tilde{t}}} \left(|\mathcal{M}_a|^2 + |\mathcal{M}_b|^2\right) &= \frac{1}{256\pi^3}\left(\frac{M_{\tilde{N}_a}}{\Gamma_{\tilde{N}_a}}\right) \left(\int\frac{dPS_2}{2M_{\tilde{t}}} \right) \\\nonumber &\times \left(|g_L^{l1}|^2 + |g_R^{l1}|^2 \right) \left(|g_L^{la}|^2 + |g_R^{la}|^2 \right)\\\nonumber &\times \left( \left(M_{\tilde{t}}^2-M_{\tilde{N}_a}^2-M_t^2 \right)\left(|g_L^{qa}|^2 + |g_R^{qa}|^2 \right) + 4M_t M_{\tilde{N}_a}~\textrm{Re}\left(g_L^{qa} g_R^{qa*}\right)\right) \\ \nonumber &\times \int_0^{1-\mu_1} dx \frac{x^2(1-x-\mu_1)^2}{1-x}\frac{1}{(1-x-\mu_{\tilde{l}})^2} 
\end{align}
and
\begin{align}
\label{eqn:IntTerms}
\int \frac{dPS_4}{2M_{\tilde{t}}} \left(2\textrm{Re}(\mathcal{M}_a\mathcal{M}_b^*) \right) &= \frac{1}{256\pi^3}\left(\frac{M_{\tilde{N}_a}}{\Gamma_{\tilde{N}_a}}\right) \left(\int\frac{dPS_2}{2M_{\tilde{t}}} \right) \\\nonumber &\times \left(\left(M_{\tilde{t}}^2-M_{\tilde{N}_a}^2-M_t^2 \right) \left(|g_L^{qa}|^2 + |g_R^{qa}|^2 \right) -  4M_t M_{\tilde{N}_a}~\textrm{Re}\left(g_L^{qa} g_R^{qa*}\right)   \right) \\\nonumber &\times \int dx_+ dx_- \frac{1}{(1-x_+-\mu_{\tilde l})(1-x_--\mu_{\tilde l})} \\\nonumber &\times \left( \sqrt{\mu_1}\left(\mu_1+x_++x_- - 1\right) \textrm{Re}\left(g_L^{l1}g_L^{l1} g_L^{la*}g_L^{la*} + g_R^{l1}g_R^{l1} g_R^{la*}g_R^{la*} \right)   \right. \\\nonumber &~~~~ -\left. 2(1-x_+-x_-+x_+x_- - \mu_1) \textrm{Re}\left(g_L^{l1}g_L^{l1} g_R^{la*}g_R^{la*}\right) \right).
\end{align}

The entire expression is relativistically invariant, except for the limits in Eq. (\ref{eqn:N+N-}) used to derive
Eq. (\ref{eqn:Difference}), which are computed in the rest frame of $\tilde{N}_a$. However, since $N_+$ ($N_-$) involves an integration over the entire upper (lower) hemisphere, these expressions are still invariant under boosts in the stop's direction which do not flip the direction of the top. In particular, the asymmetry parameter is unmodified when boosting to the rest frame of the stop. This is an important fact. It implies that the signal one constructs in the lab is only degraded by one's ignorance of the initial boost of the stop in the lab frame.

In the case of an off-shell slepton the asymmetry variable $\eta$ is an $\mathcal{O}(1)$ number. The exact expression is given by the ratio of Eq. (\ref{eqn:Difference}) to Eq. (\ref{eqn:Total}). There are several limiting cases where the final result is extremely simple. In particular, in the case where $\tilde{N}_a$ is a pure wino and $\tilde{N}_1$ is a pure bino,  assuming there are no strong kinematical suppressions and $\mu_{\tilde{l}} \gg 1$, the expression simplifies to
\begin{align}
\label{eqn:SimpleEta}
\eta = \frac{\sqrt{\mu_1}}{2}\left(\frac{F(\mu_1)}{G_1(\mu_1) + G_2(\mu_1) \cos(2\Delta\varphi)}\right) ~\sin(2\Delta\varphi),
\end{align}
where we expressed the complex couplings as $g_L^{la} = |g_L^{la}| ~\exp(i\varphi_L^{la})$ and
\begin{equation}
\Delta \varphi = \varphi_L^{l1}-\varphi_L^{la}.
\end{equation}
In the approximation of ignoring slepton mixings this is just the difference between the original
phases of the neutralino masses before they were absorbed into the couplings.
The kinematic functions in Eq. (\ref{eqn:SimpleEta}) are given by
\begin{align*}
F(\mu_1) &= \int dx_+ dx_- \Bigl( \left(1 -\mu_1-x_+-x_- +x_+x_-\right)\left(x_++x_-+\mu_1 -1\right) \Bigr)^{1/2} \\
G_1(\mu_1) &= \frac{1}{6}\left( 1-8\mu_1 + 8\mu_1^3 - \mu_1^4\right) - \mu_1^2 \log(\mu_1^2) \\
G_2(\mu_1) &= \frac{\sqrt{\mu_1}}{6} \Bigl((1-\mu_1)(1+10\mu_1+\mu_1^2) + 6\mu_1(1+\mu_1) \log(\mu_1) \Bigr)
\end{align*}

\begin{figure}
\begin{center}
\vspace*{.36cm}
\includegraphics[scale=0.4]{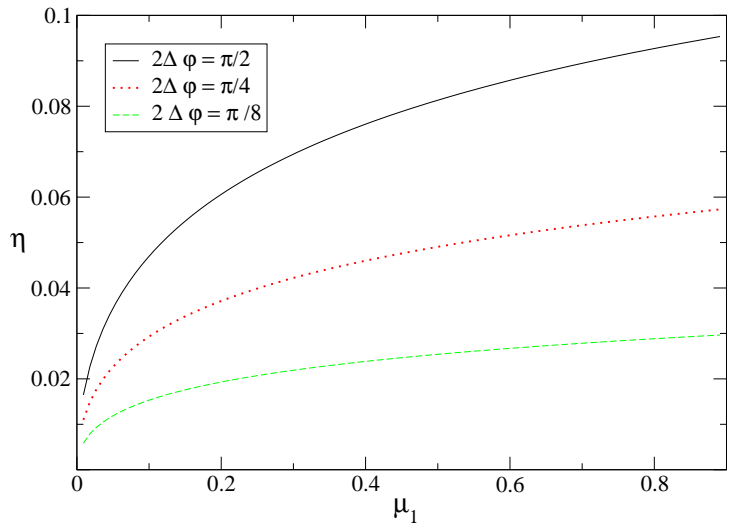}
\end{center}
\caption{The asymmetry parameter $\eta$ is plotted against $\mu_1 = M_{\tilde{N}_1}^2/M_{\tilde{N}_a}^2$ for several choices of the CP-phase $\Delta \varphi$. As expected, $\eta$ diminishes as the CP-phase decreases.} 
\label{fig:etaVsmu1}
\end{figure}

In Fig. \ref{fig:etaVsmu1}  we plot the asymmetry parameter $\eta$ vs. $\mu_1$ for several choice of $\Delta \varphi$. As claimed above, when the slepton is off-shell the asymmetry $\eta$ can be very large, proportional to the phase times an $\mathcal{O}(1)$ number as shown in Eq. (\ref{eqn:SimpleEta}). For example, if we take $M_{\tilde{N}_1}/M_{\tilde{N}_a} \gtrsim 0.7$ we find that, ignoring experimental limitations, the number of events needed to make a determination of CP-violation in this cascade decay is approximately
\begin{equation}
N = \frac{1}{\eta_{th}^2} \sim \frac{100}{\sin^2(2\Delta\varphi)}.
\end{equation}

In Table \ref{tbl:rate} we present the $\tilde{t}_L\tilde{t}_L^c$ production cross-section for several choices of the stop mass. We also show the actual number of $t~\ell^+~\ell^-$ events, taking into account the branching ratio for the reaction $\tilde{t} \rightarrow t + \tilde{N}_a \rightarrow t + l^+ + l^- + \tilde{N}_1$, for an integrated luminosity of $\mathcal{L} = 300 ~fb^{-1}$ and for a possible upgrade with $\mathcal{L} = 1~ab^{-1}$. For a stop mass below $800\GeV$ the prospects for such a measurement look promising. 
\begin{table}[h]
\begin{center}
\begin{tabular}{|c|c|c|}
\hline
$M_{\tilde{t}_L}$ & $\sigma~(fb)$ &\parbox{4.2cm}{\center{$N[t\ell^+ \ell^-]$ \\ $\mathcal{L} = 300 fb^{-1}$ (1 $ab^{-1}$)}} \\ \hline 
$500\GeV$ & $300$ &  $7300 ~~(24000)$\\
$800\GeV$ & $20$ &  $560 ~~(1800)$ \\
$1\TeV$ &  $4$ &  $120 ~~(400)$ \\
$1.2\TeV$ & $1$ & $30 ~~(100)$  \\
\hline
\end{tabular}
\caption{The production cross-section for $\tilde{t}_L{\tilde{t}}_L^c$ is shown in the middle column. The branching ratio for the reaction  $\tilde{t} \rightarrow t + \tilde{N}_a \rightarrow t + l^+ + l^- + \tilde{N}_1 $ was calculated using $M_{\tilde{l}} = 300\GeV$, $M_{\tilde N_2}=140\GeV$,  $M_{\tilde N_1}=100\GeV$, and assuming that the gluino and squarks are sufficiently heavy to have little effect.  (Under these assumptions and wino/bino dominated $\tilde N_{2,1}$ the branching ratio
for $\tilde t \rightarrow t  \tilde N_2$ is slightly less then $1/3$ because of the top's mass, and those for $ \tilde N_2\rightarrow e^+ e^- \tilde N_1$ or $\mu^+ \mu^- \tilde N_1$ are about $1/6$ each.)
The number of $t~ \ell^+~\ell^-$ events is then presented in the last column for two different integrated luminosities.
The effective number of events is doubled if one combines the $t \ell^+ \ell^-$ and $t^c \ell^- \ell^+ $asymmetries.}
\label{tbl:rate}
\end{center}
\end{table}

In the next section we take into account the experimental difficulties in making such a determination. We propose ways of overcoming these limitations and try to evaluate the corresponding reduction in signal sensitivity. 

\section{Experimental limitations}
\label{sec:ExpLim}

In this section we consider the degradation of the signal due to experimental limitations. First, we address an issue already mentioned above, namely that the triple product is measured in the lab frame, and if the LSP escapes detection there is no way to reconstruct the rest frame of the stop. In other words, the asymmetry parameter $\eta$ was computed in the neutralino frame, which cannot be reconstructed. 

Let's imagine an event where the momenta are such that in the neutralino's rest frame we have,
\begin{equation*}
\vec{p}_{t} \cdot (\vec{p}_{l^+}\times\vec{p}_{l^-}) = p_{t}^z \left(p_{l^+}^x p_{l^-}^y - p_{l^+}^y p_{l^-}^x\right) > 0
\end{equation*}
This is a contribution to $N_+$. This quantity is still positive even in the stop's rest frame since the boost is only along the $z$-axis and it cannot flip the direction of the top momenta. If the stop was produced at rest in the lab frame the signal would be unaltered. However, the stop itself is in general boosted with respect to the lab frame. An arbitrary boost can turn this contribution to $N_+$ in the stop's frame into a contribution to $N_-$ in the lab frame. It can either flip the sign of $\vec p_{t}$ or it can change the transverse orientation of $\vec p_{l^+}$ with respect to $\vec p_{l^-}$.  

\begin{figure}
\begin{center}
\includegraphics[scale=0.4]{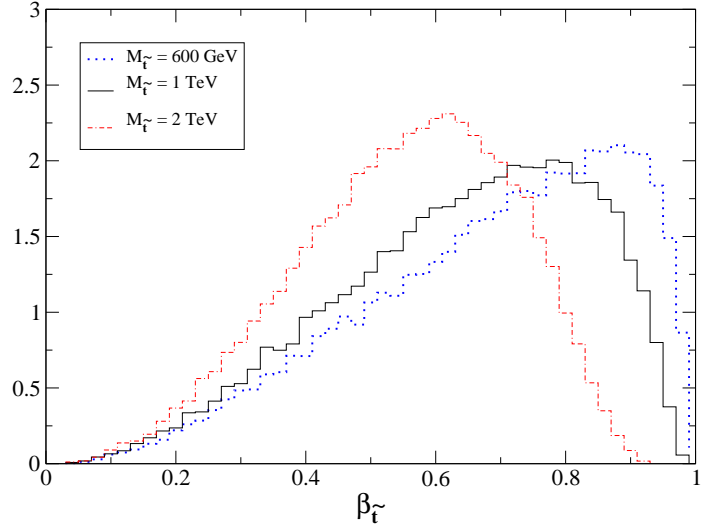}
\end{center}
\caption{The stop velocity distribution in the lab frame for several choices of stop's mass. All distributions are normalized to unit area. As the stop's mass increases the distributions peak at a lower $\beta$, in agreement with expectations based on the PDF.}
\label{fig:stop_beta}
\end{figure}

The stop - anti-stop pair is produced mainly via gluon fusion and so, owing to the gluon distribution function, the stops are produced very close to threshold. However, the overall center of mass can be quite boosted with respect to the lab. Therefore, while the stop has very little transverse momenta, it does carry a non-negligible momentum along the beam direction. In Fig. \ref{fig:stop_beta}  we used Pythia \cite{Sjostrand:2006za} to produce a plot of the distribution of stop longitudinal velocity in the lab frame for several choices of stop mass. 

Therefore, we must account for the possible flip of $N_+$ into $N_-$ (and vice-versa) due to the initial boost of the stop. If we denote by $w$ the probability  flip then the asymmetry parameter $\eta$ in the lab frame is given by
\begin{equation}
\eta_{lab} = \frac{N_+^{(lab)} - N_-^{(lab)}}{N_+^{(lab)} + N_-^{(lab)} }= \mathcal{D}\left(  \frac{N_+^{(\tilde{N})} - N_-^{(\tilde{N})}}{N_+^{(\tilde{N})} + N_-^{(\tilde{N})} }\right),
\end{equation} 
where the dilution factor $\mathcal{D}$ is simply
\begin{equation}
\mathcal{D} = 1- 2 w.
\end{equation}
 
\begin{figure}
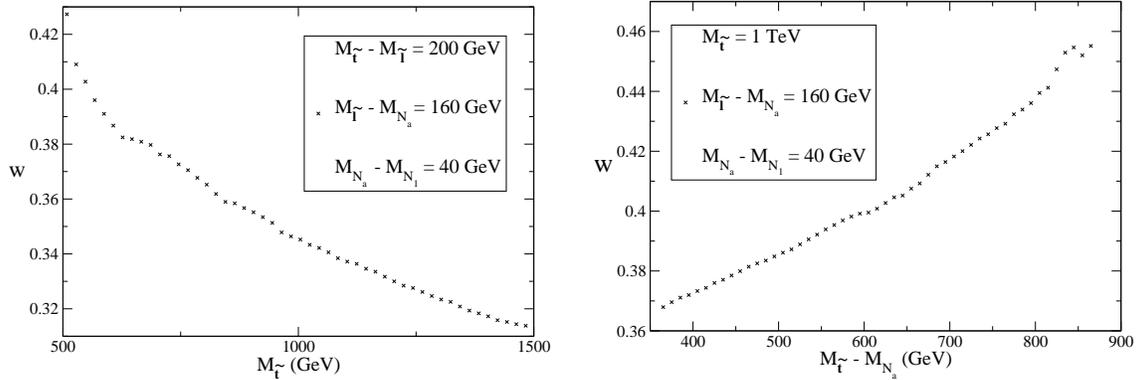

\begin{center}
\includegraphics[scale=0.29]{flipVsMstop.eps} \hspace{3mm}
\includegraphics[scale=0.29]{flipVsDM.eps}
\end{center}
\caption{In the left pane the probability $w$ of $N_+$ in the neutralino's rest frame to flip into $N_-$ in the lab frame (or $N_-$ into $N_+$) is plotted as a function of the stop mass keeping $m_{\tilde{t}} - m_{\tilde{N}_a}$ fixed. In the right pane, $w$ is plotted as a function of the mass difference, keeping $m_{\tilde{t}}$ fixed.} 
\label{fig:dilution_fac}
\end{figure}

One might expect that $w\rightarrow 0$ as the stop's mass increases since the initial boost is diminished. However, this limit holds true only if the mass difference $m_{\tilde{t}} - m_{\tilde{N}_a}$ remains fixed. When this mass difference increases, $w$ increases as well. To understand this point, notice that when the difference $m_{\tilde{t}} - m_{\tilde{N}_a}$ increases, all the momenta defining $N_+$ are on average increased. So, while it is true that it becomes harder to change the sign of $\vec p_{t}$, it is easier to change the orientation of $\vec p_{l^+}$ with respect to $\vec p_{l^-}$. In Fig. \ref{fig:dilution_fac}  we plot the probability for a flip, $w$, as a function of the stop's mass (keeping $m_{\tilde{t}} - m_{\tilde{N}_a}$ fixed) as well as a function of the mass difference itself (keeping $m_{\tilde{t}}$ fixed). 

From Fig. \ref{fig:dilution_fac} it is clear that the dilution factor does not present a very serious problem. Unless there is a very large splitting between the stop and neutralino masses the probability of flip is about $w\gtrsim 0.33$. This will translate into a dilution factor of about $\mathcal{D}^2 \lesssim 0.1$ which represents an increase in the number of events needed of about an order of magnitude. 

There are several related issues concerning the efficiency of identifying the top or anti-top, determining its 
charge\footnote{Averaging over both the top and anti-top processes, which correspond to 
$\langle{ \cal T} \rangle -\langle{\overline{\cal T}}\rangle$ in Section \ref{sec:GeneralDiscussion}, cancels the CP-violating effects and just leaves the
strong phase contribution.} 
and momentum, determining whether the $l^+ l^-$ from the decay is associated with the initial 
$\tilde t$ or the $\tilde t^c$ (assuming they are pair produced), separating the cascade leptons from others
in the process, etc.  These questions, and how they affect the number of identified events,
will require an extensive and careful numerical simulation. While we do not attempt such a study here but leave it for future research, we would like to make a number of comments relevant for such a study. A favorable situation is that in
which $\tilde N_1$ and $\tilde{N}_2$ are dominantly bino and wino, respectively,  and the gluino is heavy. Then the dominant decays
of the $\tilde t_L$ should be into $b~ \tilde C^+$ and $t~ \tilde N_2$, with relative rates $\sim 2 : 1$. When one
of the stops decays to a chargino and the other cascades, it may optimistically be possible to
identify the $t$ or $t^c$ and determine its charge by tagging on the charge of the lepton from
the chargino decay (especially if it is of a different flavor from the dilepton). 
 In this case, one can combine the asymmetries from $\tilde t$ and $\tilde t^c$
decays; i.e., in analogy to Eq.  (\ref{eqn:eta_def}) define 
\begin{equation}
\label{eqn:etasum_def}
\eta_{sum} = \frac{N_+(\tilde t)+N_+(\tilde t^c) - N_-(\tilde t)-N_-(\tilde t^c) }{N_+(\tilde t)+N_+(\tilde t^c)  + N_-(\tilde t)+N_-(\tilde t^c) },
\end{equation}
where the $\theta$ angles for the $\tilde t$ and $\tilde t^c$ events are defined in the caption to Fig.  \ref{fig:z-plane-angle}.
It is easy to show that the theoretical expectation for $\eta_{sum}$ is the same as for $\eta$ (except for removing the strong
phase term), while effectively doubling the number of available events, i.e., the required $({\mathcal{D}^2 \eta_{th}^2})^{-1}$
is the total number of $t ~\ell^+\ell^-$ and $t^c \ell^-\ell^+$ cascades.

Events in which  the associated top
decays semi-leptonically provide  another handle on its identification and charge,
but at the expense of possible additional confusion about which  lepton is from the top, and missing
the neutrino momentum,  which renders the reconstruction of the top's momentum impossible. None of the observables above really require the top's momentum, but only its direction in the lab frame. If the top is highly boosted, it may not be necessary to fully reconstruct its momentum to infer its direction in the lab frame. More generally, the reduction in signal due to an imprecise determination of the top's direction is a detailed numerical question involving an event simulator which we leave for future investigation. 

There may also be a non-negligible number of events in which the $\tilde t$ decays to $t~\tilde N_2$
and the $\tilde t^c$ to $t^c \tilde N_1$, or vice versa. Assuming that one determines the charge
of $t$ or $t^c$ by its leptonic decay, there is still the ambiguity of whether the dilepton is correlated
with the top or anti-top. One possibility is to simply include the wrong pairing (calculating $\cal T$ ($\overline{\cal T}$)
for the
pairing with the $t$ ($t^c$)), and argue that on average it does not contribute to any CP violating observable. 
This effectively sums the asymmetries from  the $t~ \ell^+\ell^-$ and $t^c \ell^-\ell^+$ cascades, but without gaining the factor two
statistical advantage discussed above in the case of one chargino decay.
It is possible that the combinatorics can be resolved with a more sophisticated analysis involving isolation cuts, energy cuts, etc. For example, if the anti-stop decays directly into an anti-top and the LSP, the anti-top is on average more energetic than the top coming from the other branch. Ordering the jets according to their $p_T$ may reduce the combinatorics and help identify the correct pairing. Again, such schemes will require a careful numerical simulation.

Another possibility is to use bottom quarks instead of tops, i.e., consider the reaction $\tilde{b} \rightarrow b + \tilde{N}_2 \rightarrow b + l^+ + l^- + \tilde{N}_1$.  In this case one could reconstruct the full $b$ momentum. However, 
the efficiency of directly determining the $b$ charge is low.
These issues could be resolved if the opposite side $\tilde b$ or $\tilde b^c$ decays to $t~\tilde C^-$
or $t^c~\tilde C^+$, but it may be difficult to know whether one began with stops or sbottoms if they are close in mass. For example, if one observed $t~ b^c l^{\prime -} \ell^+\ell^-$, the $\ell^+\ell^-$ could be associated either with $\tilde{t} \rightarrow t ~\ell^+~\ell^-$ or with $\tilde{b}^c\rightarrow b^c~\ell^-~\ell^+$. The asymmetries expected from each possibility would be the same for a wino-dominated $\tilde{N}_2$. Similar to the discussion of $t^c\tilde{N}_1$ above, one could count each event twice, once
for each possibility, assuming that the wrong pairing does not contribute to the asymmetry. The required sum of  identified $t$ and $b^c$ cascades
is twice the expression in Eq. (\ref{eqn:nstat}), but this number could include the
$t^c$ and $b$ cascades if the $t^c b \ell^{\prime +} \ell^-\ell^+$ events are combined appropriately.

A third possibility is to use the asymmetry between quarks and anti-quarks in the parton distribution function (PDF) of the proton. When considering stop pair-production the dominant mode is gluon fusion since there are practically no tops in the proton PDF. In this case, stops and anti-stops are produced in equal amounts. However, when considering the production of $\tilde{u}$'s and $\tilde{d}$'s there are many more relevant channels. The valence quarks play a significant role and associated production ($g+q\rightarrow \tilde{g}+\tilde{q}$) can dominate. Since valence quarks in a proton-proton collider are mostly quarks and not anti-quarks, it is considerably more likely to produce a squark than an anti-squark. Therefore, in effect, we know that we are observing the reaction and not its CP conjugate. There are several problems with such an approach. First, we must include the contribution from all partons (mostly $u$'s and $d$'s). This is in principle easy to incorporate into the present calculation and amounts to a trivial addition of the different contributions. The more serious problem is the existence of multijets in the event and a reduction of the signal due to combinatorics.

There is also the problem of the strong phases discussed in Section  \ref{sec:GeneralDiscussion},
which in principle contribute to the expectation value of the triple-product.
One source involves the exchange of a photon between the two leptons in the cascade. However, this
is of $\mathcal{O}(\alpha/\pi)$, and not competitive unless the CP-violating phases are small. Another is associated with the
phases in the slepton propagators from their finite width. This leads to a non-trivial effect in the on-shell
case for fixed lepton energies, but vanishes when integrated over their energies, as commented in the appendix.
It is negligible in the more interesting off-shell case. In general, the strong phase effect can
be eliminated by combining the asymmetries for $\tilde t$ and $\tilde t^c$, as in Eq. (\ref{eqn:sum_value}) or (\ref{eqn:etasum_def}).
However, if one relies on the asymmetry between quarks and anti-quarks in the PDF, one cannot form such a combination. In this case, one must rely on the theoretical estimate that the strong phases effects are small. 

\section{Conclusions}
\label{sec:conclusions}
The triple product and the related asymmetry parameter observable presented in this paper are sensitive to CP-phases in the cascade decay of stops, $\tilde{t} \rightarrow t + l^+ + l^- + \tilde{N}_1$. The phase combination that appears in this reaction involves the phase difference between the wino-slepton-lepton and bino-slepton-lepton 
(more generally, $\tilde N_2$ and $\tilde N_1$) couplings. As pointed in the introduction, this phase combination is not bounded directly by EDM experiments since it does not require a higgsino insertion and is therefore independent of the $\mu$-parameter's phase.

In the case of an on-shell cascade decay, the signal is too small to be observable, requiring more than $10^5$ events to reach experimental sensitivity. However, if the spectrum is such that the reaction proceeds via an off-shell slepton, the signal is greatly enhanced. We find that about $(10^2-10^3)/\sin^2(2\Delta\varphi)$ events are needed to constrain the CP-phase $\Delta\varphi$. This number may improve dramatically if some experimental difficulties discussed in the text are resolved or may increase if these turn out to be more severe.  At any rate, this number is low enough to be taken seriously as a viable observable for probing some of the MSSM's CP-phases at the LHC. Rough estimates for the required number of events are given in Table \ref{tbl:rate}. For large CP phases the effect may be observable for stop mass as large as $800\GeV$ prior to a luminosity upgrade and even higher thereafter. A possible luminosity upgrade and a favorable spectrum will place the signal well within the experimental sensitivity and help probe a combination of the phases which is currently inaccessible via the EDM experiments.

Our goal has been to illustrate the general possibility and point out the difficulties, not to examine
the full parameter space. A more systematic study, including a full numerical simulation of the events and
detector performances, would be very useful. Similar effects might also be observable in other channels which may cover different regions of parameter space. 

\textbf{Acknowledgments}: We would like to thank G. Kane and T. Han for useful discussions in the early stages of this project. The work of L.W. and I.Y. is supported by the National Science Foundation under Grant No. 0243680 and the Department of Energy under grant \# DE-FG02-90ER40542. P.L is supported by the Friends of the IAS and by the NSF grant PHY-0503584. The work of G.P. was supported in part by the Department of Energy \# DE-FG02-90ER40542 and by the United States-Israel Bi-national Science Foundation grant \# 2002272. 
Any opinions, findings, and conclusions or recommendations expressed in this material are those of the author(s) and do not necessarily reflect the views of the National Science Foundation. 

\appendix
\renewcommand{\theequation}{A-\arabic{equation}}
\setcounter{equation}{0}

\section{A derivation of the asymmetry parameter $\eta$ in the neutralino's rest frame}
\label{app:EtaComp}
In this appendix we give the details of the calculation of the
asymmetry $\eta$ in the neutralino's rest frame
(defined in Eq.(\ref{eqn:eta_def})). A computation of the
expectation value $\langle \epsilon_{\mu\nu\alpha\beta}\; p_{\tilde t}^\mu
p_t^\nu p_{l^+}^\alpha p_{l^-}^\beta~\rangle$ is a straightforward
modification of the derivation below.

The differential decay width for the reaction is
\begin{equation}
 d\Gamma = \frac{\sum_{spin}|\mathcal{M}|^2}{2M_{\tilde t}} dPS_4,
\end{equation}
where $\mathcal{M}$ is the invariant amplitude. The Feynman rule for the sfermion-fermion-neutralino
coupling is
\begin{equation}
\vcenter{\includegraphics[scale=0.75]{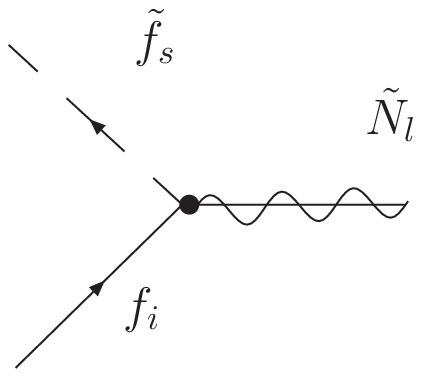}}
\hspace{-28em} \quad = \quad\quad\quad i\left(G^{f_L}_{isl}\,P_L+G^{f_R}_{isl}\,P_R\right).
\end{equation}
Our notation follows that of \cite{Drees:2004jm}, where the explicit
expressions for the $G$'s in terms of the mixing matrices can be
found.

The invariant amplitude consists of two parts, corresponding to the
Feynman diagrams of Fig. \ref{fig:Mab}. (There are two diagrams contributing to
the process as a result of the Majorana nature of the neutralinos).
Ignoring the masses of the external leptons and allowing them to have different
flavors the two parts are
\begin{eqnarray}
\label{eqn:MaMb}
i\mathcal{M}_a &=& i \bar{u}(p_t) \left(G^{t_R*}_{t\tilde{t}a}P_L +
G^{t_L*}_{t\tilde{t}a}P_R \right) \left(\frac{-\slashed{q} +
M_{\tilde{N}_a}}{q^2-M_{\tilde{N}_a}^2} \right) \left(G^{\,l_R*}_{jka}P_L +
G^{\,l_L*}_{jka}P_R \right) v(p_{l^-_j}) \nonumber\\&&
\left(\frac{1}{k_1^2 - M_{\tilde{l}_k}^2}\right) \bar{u}(p_{\tilde N_1})
\left(G^{\,l_L}_{ik1}P_L + G^{\,l_R}_{ik1}P_R \right) v(p_{l^+_i})
\nonumber\\
i\mathcal{M}_b &=& -i \bar{u}(p_t)
\left(G^{t_R*}_{t\tilde{t}b}P_L + G^{t_L*}_{t\tilde{t}b}P_R \right)
\left(\frac{-\slashed{q} + M_{\tilde{N}_b}}{q^2-M_{\tilde{N}_b}^2} \right)
\left(G^{\,l_L}_{ilb}P_L + G^{\,l_R}_{ilb}P_R \right)
v(p_{l^+_i})\nonumber \\&& \left(\frac{1}{k_2^2 -
M_{\tilde{l}_l}^2}\right) \bar{u}(p_{\tilde N_1})
\left(G^{\,l_R*}_{jl1}P_L + G^{\,l_L*}_{jl1}P_R \right) v(p_{l^-_j}),
\end{eqnarray}
where $q=p_{\tilde{t}}-p_t$, $k_1=p_{l^+_i}+p_{\tilde N_1}$,
$k_2=p_{l^-_j}+p_{\tilde N_1}$, and the masses are taken to be real
(i.e., all the complex phases are absorbed into the couplings). In
deriving \ref{eqn:MaMb} we have used some of the identities of
Appendix C of \cite{Chung:2003fi}.

The asymmetry $\eta$ depends on the terms proportional to the
Levi-Civita tensor $\epsilon_{\mu\nu\alpha\beta}$. Terms in the
reaction containing a non-vanishing $\epsilon_{\mu\nu\alpha\beta}$ can
only come from the interference terms and must contain 4 independent
vectors. We find

\begin{eqnarray}
&&2\,\textrm{Re}\sum_{\rm spin} \mathcal{M}_a \mathcal{M}_b^*
\supset  4\, \textrm{Im}\,\left(a_R - a_L\right)
\epsilon_{\mu\nu\alpha\beta}~ p_{\tilde{t}}^\mu ~p_t^\nu~
p_{l^+_i}^\alpha~ p_{l^-_j}^\beta~\\ &&\times\left[
\frac{1}{q^2-M_{\tilde{N}_a}^2-i\Gamma_{\tilde{N}_a} M_{\tilde{N}_a}}\cdot
\frac{1}{q^2-M_{\tilde{N}_b}^2+i\Gamma_{\tilde{N}_b} M_{\tilde{N}_b}}\cdot
\frac{1}{k_1^2-M_{\tilde{l}_k}^2-i\Gamma_{\tilde{l}_k}
M_{\tilde{l}_k}}\cdot
\frac{1}{k_2^2-M_{\tilde{l}_l}^2+i\Gamma_{\tilde{l}_l}
M_{\tilde{l}_l}} + {\rm c.c} \right],\nonumber
\end{eqnarray}
where $a_R$ is given by
\begin{eqnarray}
a_R
&=&-q^2\left(G^{t_L*}_{t\tilde{t}a}G^{t_L}_{t\tilde{t}b}G^{\,l_R*}_{jka}
G^{\,l_L*}_{ilb}G^{\,l_L}_{ik1}G^{\,l_R}_{jl1}\right)\\ \nonumber
&+&M_{\tilde{N}_a}M_{\tilde
N_1}\left(G^{t_R*}_{t\tilde{t}a}G^{t_R}_{t\tilde{t}b}G^{\,l_R*}_{jka}G^{\,l_R*}_{ilb}
G^{\,l_R}_{ik1}G^{\,l_R}_{jl1}\right)\\ \nonumber &+& M_{\tilde{N}_b}M_{\tilde
N_1}\left(G^{t_R*}_{t\tilde{t}a}G^{t_R}_{t\tilde{t}b}G^{\,l_L*}_{jka}G^{\,l_L*}_{ilb}
G^{\,l_L}_{ik1}G^{\,l_L}_{jl1}\right)\\ \nonumber
&+&M_{\tilde{N}_a}M_{\tilde{N}_b}\left(G^{t_R*}_{t\tilde{t}a}G^{t_R}_{t\tilde{t}b}G^{\,l_R*}_{jka}G^{\,l_L*}_{ilb}
G^{\,l_L}_{ik1}G^{\,l_R}_{jl1}\right),
\end{eqnarray}
and $a_L$ is simply given by $a_R$ with $L\leftrightarrow R$.
In principle, there is also a term proportional to $\textrm{Re}\ (a_R - a_L)$ from
the finite width part of the slepton propagators (see Section \ref{sec:conclusions}). In the flavor diagonal case this
term is proportional to the difference between the $l^+$ and $l^-$ energies, and
vanishes when integrated over phase space.

If we neglect all the off-diagonal mixing matrix elements, and take
the neutralino $\tilde{N}_a=\tilde{N}_b$ to be on-shell
($q^2=M_{\tilde{N}_a}^2$), the expression simplifies to

\begin{align}
\label{eqn:Im_diag}
\textrm{Im}\,(a_R-a_L) &= 2M_{\tilde{N}_a}^2\left(|g^{qa}_R|^2 - |g^{qa}_L|^2
\right) \textrm{Im}\,\left( g^{la*}_Rg^{la*}_L g^{l1}_Rg^{l1}_L
\right)\\\nonumber & + M_{\tilde{N}_a} M_{\tilde N_1} \left(|g^{qa}_R|^2 -
|g^{qa}_L|^2 \right) ~ \textrm{Im}\,\left[\left(g^{la*}_R \right)^2
\left(g_R^{l1}\right)^2 +\left(g^{la*}_L \right)^2
\left(g_L^{l1}\right)^2\right],
\end{align}
where we have defined
\begin{equation}
G^{t_R}_{t\tilde{t}a}\equiv g^{qa}_R,\quad G^{\,l_R}_{ika}\equiv g^{la}_R,\quad G^{\,l_R}_{ik1}\equiv
g_R^{l1},
\end{equation}
and similarly for the left handed couplings. Writing each coupling constant as $g=|g|e^{i\varphi}$ (\ref{eqn:Im_diag}) can be expressed in terms of the complex phases $i\varphi$ as

\begin{align}
\textrm{Im}\,(a_R-a_L) &= 2M_{\tilde{N}_a}^2\left(|g_R^{qa} |^2 - |g_L^{qa} |^2
\right) | g_R^{la}|\, | g_L^{la}|\,  |g_R^{l1}|\,  |g_L^{l1}|
\sin\left( \varphi_R^{l1} + \varphi_L^{l1} -\varphi_R^{la} -
\varphi_L^{la})\right)   \\\nonumber 
& +  M_{\tilde{N}_a} M_{\tilde N_1} \left(|g_R^{qa} |^2 - |g_L^{qa} |^2
\right) \left[|g_R^{la}|^2\,
|g_R^{l1}|^2\sin\left(2(\varphi_R^{l1} -\varphi_R^{la})\right) +
(R\rightarrow L) \right].
\end{align}

The event geometry in the neutralino's rest frame is depicted in
Fig. \ref{fig:z-plane-angle}. The incoming stop and
outgoing top define a $z$-axis with the top pointing in the positive
direction. This $z$-axis is in an arbitrary orientation with respect to
the lab frame's beam pipe-line axis (since the stop is a scalar its
decay is isotropic). Momentum conservation in the neutralino's rest
frame forces the di-lepton and the LSP to lie in the same plane. In other
words, the di-lepton defines an orthogonal to the plane,
\begin{equation}
\hat{n} = \hat{p}_{l^+} \times \hat{p}_{l^-},
\end{equation}
where $\hat{n}$ itself is oriented with respect to the $z$-axis,
\begin{equation}
\hat{n}\cdot \hat{p}_t = \cos\theta.
\end{equation}
We say that $\hat{n}$ is in the upper-hemisphere ($N_+$) if
$\cos\theta > 0$ or the lower-hemisphere ($N_-$) if $\cos\theta < 0$.

A non-zero expectation value for $\vec{p}_t \cdot
(\vec{p}_{l^+}\times\vec{p}_{l^-})$ translates into a non-zero
expectation value for $N_+ - N_-$. As far as this difference  is
concerned the only relevant part of the amplitude is the one
involving the $\epsilon_{\mu\nu\alpha\beta}$ piece. We are left with evaluating
the integral
\begin{eqnarray}
N_+&\propto& \int_{\theta=0}^{\pi/2} dPS_4 \left[ 4\, \textrm{Im}\,\left(a_R -
a_L\right) \epsilon_{\mu\nu\alpha\beta}~p_{\tilde{t}}^\mu ~p_t^\nu~
p_{l^+}^\alpha~ p_{l^-}^\beta\right] \\\nonumber &\times&
\frac{1}{(q^2-M_{\tilde{N}_a}^2)^2+\Gamma_{\tilde{N}_a}^2M_{\tilde{N}_a}^2}\, {\rm Re} \left(
\frac{1}{k_1^2-M_{\tilde{l}}^2-i\Gamma_{\tilde{l}} M_{\tilde{l}}}\cdot
\frac{1}{k_2^2-M_{\tilde{l}}^2+i\Gamma_{\tilde{l}} M_{\tilde{l}}}
\right).
\end{eqnarray}
A similar expression holds for $N_-$ only with the limits
on the integrals being $(\pi/2, \pi)$.
In the neutralino's rest frame $\vec{p}_{\tilde t} = \vec{p}_t$, and therefore
\begin{align}
\epsilon_{\mu\nu\alpha\beta}~ p_{\tilde t}^\mu ~p^\nu_t~ p_{l^+}^\alpha~ p_{l^-}^\beta
&= E_{\tilde t}\, \vec{p}_t \cdot \left( \vec{p}_{l^+}\times \vec{p}_{l^-}\right) - E_t\,
\vec{p}_{\tilde t}  \cdot \left( \vec{p}_{l^+}\times \vec{p}_{l_-}\right)  \\
\nonumber &= M_{\tilde{N}_a}\,  \vec{p}_t \cdot \left( \vec{p}_{l^+}\times
\vec{p}_{l^-}\right) \\ \nonumber &= M_{\tilde{N}_a}\, |p_t|\, |p_{l^+}|\, |p_{l^-}| \cos\theta
\sin\phi,
\end{align}
where $\cos\theta$ is defined as above and $\phi$ is the angle
between $\vec{p}_{l^+}$ and $\vec{p}_{l^-}$.

The four-body phase space can be written as
\begin{align}
dPS_4(p_{\tilde t} \rightarrow p_t+p_{l^+}+p_{l^-}+p_{\tilde N_1}) &= dPS_2(p_{\tilde t}\rightarrow p_t +
q)~ \frac{dq^2}{2\pi} \\  \nonumber &\times dPS_3(q\rightarrow
p_{l^+}+p_{l^-}+p_{\tilde N_1}),
\end{align}
where the 2-body phase space integral is given by
\begin{equation}
 dPS_2(p_{\tilde t}\rightarrow p_t +q) = (2\pi)^4\delta^{(4)}(p_{\tilde t}- p_t -q) \frac{d^3p_t}{(2\pi)^32E_t}\frac{d^3q}{(2\pi)^32E_q}
\end{equation}
The 3-body phase-space integral can be written in terms of
dimensionless variables (see for example \cite{Barger:1987nn})
\begin{equation}
dPS_3(q \rightarrow p_{l^+}+p_{l^-}+p_{\tilde N_1}) = \frac{M_{\tilde{N}_a}^2}{256\pi^3} dx_+ dx_-
~d\cos\theta,
\end{equation}
where
\begin{equation}
x_+ = \frac{2 E_{l^+}}{M_{\tilde{N}_a}} \quad \text{and} \quad x_- = \frac{2
E_{l^-}}{M_{\tilde{N}_a}},
\end{equation}
and the limits of integration are
\begin{equation}
0<x_-<1-\mu_1,\quad 1-\mu_1-x_-<x_+<1-\frac{\mu_1}{1-x_-},
\end{equation}
where $\mu_1 = M_{\tilde N_1}^2/M_{\tilde{N}_a}^2$.
Despite its appearance, the integration domain is symmetric over $x_+ \leftrightarrow x_-$.

In the narrow-width approximation, the neutralino's
propagator is
\begin{equation}
\frac{1}{(q^2-M_{\tilde{N}_a}^2)^2+\Gamma_{\tilde{N}_a}^2M_{\tilde{N}_a}^2} \rightarrow
\frac{\pi}{\Gamma_{\tilde{N}_a} M_{\tilde{N}_a}} \delta (q^2 - M_{\tilde{N}_a}^2),
\end{equation}
and the $dq^2$ integration can be done trivially, setting $q^2 =
M_{\tilde{N}_a}^2$ everywhere else in the expression.
Conservation of momentum fixes the angle between $\vec{p}_{l^+}$ and
$\vec{p}_{l^-}$ to be
\begin{equation}
x_+ x_- \sin\phi =
2\left((1-\mu_1-x_+-x_- +x_+x_-)(x_++x_++\mu_1-1) \right)^{1/2}.
\end{equation}
Noting that $k_1^2 = (q-p_{l^-})^2 = M_{\tilde{N}_a}^2(1-x_-)$ and $k_2^2 =
(q-p_{l^+})^2 = M_{\tilde{N}_a}^2(1-x_+)$, the expression for $N_+$ simplifies to
\begin{eqnarray}
N_+ &\propto& \left(\int\frac{dPS_2}{2M_{\tilde t}}\right)
\frac{1}{256\pi^3}\frac{M_{\tilde{N}_a} |\vec{p}_t|}{\Gamma_{\tilde{N}_a}M_{\tilde{N}_a}}\,\textrm{Im}\,(a_R-a_L)\\  \nonumber &\times &  \int_{\cos\theta =
0}^{1}\cos\theta d\left(\cos\theta\right) \int dx_+ dx_-
\frac{x_+ x_-
\sin\phi}{\left(1-x_+-\mu_{\tilde{l}}\right)\left(1-x_--\mu_{\tilde{l}}\right)},
\end{eqnarray}
where $\mu_{\tilde l} = M_{\tilde l}^2/M_{\tilde{N}_a}^2$.

Integrating over $\cos\theta$, we arrive at the result quoted in the
text for the difference $N_+ - N_-$ for a given integrated luminosity $\mathcal{L}$,
\begin{align}
\frac{N_+ - N_-}{\mathcal{L}} &= \frac{1}{256\pi^3}\left(\frac{M_{\tilde{N}_a}}{\Gamma_{\tilde{N}_a}}\right)
\left(\int\frac{dPS_2}{2M_{\tilde t}} \right)
\left(\frac{|\vec{p}_t|}{M_{\tilde{N}_a}}\right) \textrm{Im}\,(a_R-a_L) 
\\\nonumber &\times \int dx_+ dx_-
\frac{\Bigl((1-\mu_1-x_+-x_-+x_+x_-)(x_++x_++\mu_1-1)
\Bigr)^{1/2}}
{\left(1-x_+-\mu_{\tilde{l}}\right)\left(1-x_--\mu_{\tilde{l}}\right)}.
\end{align}
For the total number of events $N_+ + N_-$ we must include the entire
amplitude
\begin{equation}
\frac{N_{\rm total}}{\mathcal{L}} = \int \frac{dPS_4}{2M_{\tilde t}} \left(|\mathcal{M}_a|^2 +
|\mathcal{M}_b|^2 + 2\textrm{Re}(\mathcal{M}_a\mathcal{M}_b^*)
\right).
\end{equation}
Following the same path outlined above it is straightforward to
arrive at the result quoted in Eqs.(\ref{eqn:notIntTerms}) and
(\ref{eqn:IntTerms}).

\bibliography{ref-triProd}
\bibliographystyle{jhep}
\end{document}